\documentclass[a4paper,11pt]{article}

\usepackage[utf8]{inputenc}
\usepackage[T1,T2A]{fontenc}
\usepackage{amsmath,amsfonts,amssymb}

\usepackage{cite}

\usepackage{graphicx}
\usepackage{wrapfig}
\usepackage{hyperref}

\usepackage[bottom]{footmisc}

\usepackage{geometry}
\newcommand{\lb}[0]{\left(}
\newcommand{\rb}[0]{\right)}
\newcommand{\lsb}{\left[}
\newcommand{\rsb}{\right]}

\begin{document}

\renewcommand*{\thefootnote}{\fnsymbol{footnote}}

\begin{center}
{\large\bf  A second Higgs near 0.5 TeV from bottom-up holographic modeling
of beyond the Standard Model strong sector}
\end{center}
\bigskip
\begin{center}
{ Sergey Afonin\footnote{E-mail: \texttt{s.afonin@spbu.ru}.}
}
\end{center}

\renewcommand*{\thefootnote}{\arabic{footnote}}
\setcounter{footnote}{0}

\begin{center}
{\small\it Saint Petersburg State University, 7/9 Universitetskaya nab.,
St.Petersburg, 199034, Russia}
\end{center}

\bigskip

\begin{abstract}
One of the simplest extensions of the Standard Model (SM) consists in adding a scalar singlet.
This second Higgs boson is able to solve several fundamental problems of SM.
Additional scalar particles arise naturally in composite Higgs scenarios in which
some confining ``strong sector'' beyond the SM drives the electroweak symmetry breaking.
The underlying strongly coupled gauge theory may be similar to QCD and could be modeled holographically.
We construct a bottom-up holographic model for description of the spectrum of composite Higgs particles.
The model is based on the holographic Soft Wall model and the Wilson confinement criterion.
The constructed model predicts the existence of a second Higgs boson with a mass of about 515 GeV.
\end{abstract}

\bigskip


\section{Introduction}

The Standard Model (SM) of particle physics postulates the scalar Higgs boson as a fundamental particle.
This theory, however, contains many input parameters and for this reason is known to be rather phenomenological.
It does not explain, for instance, the dynamical origin of the electroweak symmetry breaking and
the observed value of Higgs mass. There is an old but still attractive idea that the Higgs
boson might represent a bound state of a new, beyond the SM (BSM) strongly-interacting dynamics not much above
the weak scale (see, e.g., the review~\cite{Bellazzini:2014yua}). This could solve the SM hierarchy problem, as quantum
corrections to its mass would be saturated at the compositeness scale. Considerable
theoretical progress in constructing such dynamical models was made from the
possible holographic connection between gravity in five-dimensional curved space-times and four-dimensional
strongly-coupled gauge theories~\cite{Contino:2010rs}. The holographic approach was originally inspired by the
AdS/CFT correspondence in string theory~\cite{mald,witten,gub} but turned out to be unexpectedly
successful in various areas of physics where the problem of strong coupling arises.

A general feature of composite Higgs models is a prediction of additional Higgs-like scalars.
On the other hand, an extension of the SM by adding extra scalars is able to solve several fundamental problems.
In the simplest case, one real extra scalar is enough for this purpose. The isosinglet nature of the additional
scalar would allow to avoid various constraints from the observable data, see, e.g.,
the discussions and relevant references in~\cite{Godunov:2015nea,Falkowski:2015iwa}.
For this reason, such a minimal extension of the SM has attracted considerable attention (a short review is contained in~\cite{MartinLozano:2015vtq}):
The existence of the additional scalar can stabilize the metastable electroweak vacuum of the SM,
it can provide first order electroweak phase transition needed for electroweak baryogenesis,
it can act as a portal connecting the SM to dark matter or even be a dark matter candidate. Bounds
on the given extra Higgs boson from electroweak radiative corrections and its observation in the
resonant double Higgs production processes in the future high-luminosity LHC Runs were thoroughly analyzed
in many papers, see~\cite{Godunov:2015nea,Falkowski:2015iwa,MartinLozano:2015vtq,No:2013wsa,Barger:2014taa,Chen:2014ask} and references therein.
The detailed analysis in these papers demonstrated that a second Higgs particle could be indeed discovered at the LHC.
The experimental signatures of its production and decay were shown to depend on the mass of a new scalar particle.
In light of these expectations, prediction of mass for a hypothetical heavier Higgs boson becomes an interesting problem.

In the present work, we propose a bottom-up holographic model that predicts the mass of a second Higgs particle.
By assumption, the BSM strongly-interacting dynamics above the weak scale 246~GeV is described by some strongly coupled
field theory with the gauge group $SU(N)$ or related to $SU(N)$. The theory is analogous to QCD --- it is confining and
does not change much if the large-$N$ limit~\cite{hoof,wit} is taken. The latter property is important for applicability of a holographic
description~\cite{mald,witten,gub}. Perhaps the most successful bottom-up holographic model
describing the confining properties of QCD is the Soft Wall (SW) model~\cite{son2}. If the BSM strongly coupled theory is analogous to QCD,
the phenomenological holographic models for this BSM strongly coupled theory should look similar to the phenomenological holographic models for QCD.
We will apply to the BSM Higgs sector the recent ideas put forward in Ref.~\cite{Afonin:2021zdu} for holographic
description of the scalar mesons within a generalized SW model.

Below we give a brief overview of the scalar SW holographic model and its generalization, which is used further, discuss the proposed
in~\cite{Afonin:2021zdu} way for fixation of a free model parameter with the help of holographic Wilson loop confinement criterion
and then apply this model for a holographic description of composite Higgs sector.

\section{Generalized SW holographic model}

The SW holographic model proposed by Son et al.~\cite{son2} is defined by the action
\begin{equation}
\label{SW}
  S=\int d^4xdz\sqrt{G}e^{cz^2}\mathcal{L},
\end{equation}
where \(G=|\det{G_{MN}}|\), \(G_{MN}\) is the metric of Poincare patch (\(z>0\)) of five-dimensional anti-de Sitter (AdS\(_5\)) space
that has the line element
\begin{equation}
\label{metr}
  ds^2=G_{MN}dx^Mdx^N=\frac{R^2}{z^2}\lb\eta_{\mu\nu}dx^\mu dx^\nu-dz^2\rb.
\end{equation}
Here \(R\) is the radius of the AdS\(_5\) space and \(z\) is the fifth (called holographic) coordinate.
The function $e^{cz^2}$ represents the dilaton background, $c$ is a constant introducing the mass scale, and
$\mathcal{L}$ is the Lagrangian density of some fields in AdS\(_5\) space. By assumption, these fields are
dual on the conformal boundary of AdS\(_5\) space (situated at $z=0$) to some QCD operators. The model is defined in the probe
approximation, i.e., the metric is not backreacted by matter fields and dilaton (in some sense, this backreaction
is effectively parametrized by the form of the dilaton background~\cite{Afonin:2021cwo}).

Since the background space is curved, the model is nontrivial even in the case of free fields.
The Lagrangian density $\mathcal{L}$ is then quadratic in fields and following the prescriptions
of AdS/CFT correspondence~\cite{witten,gub} one can obtain nontrivial two-point correlation functions while
higher-order correlation functions will not appear. Remembering that the holographic approach is
formulated only in the large-$N$ limit of conformal field theories with $SU(N)$ gauge group, the absence
of higher-order correlation functions reflects the fact that
they are suppressed by positive powers of $1/N$ in confining $SU(N)$ gauge theories at $N\gg1$~\cite{wit}.

Consider the case of free scalar fields,
\begin{equation}
\label{11}
\mathcal{L}=\frac12\left(G^{MN}\partial_M\!\Phi\partial_M\!\Phi-m_5^2\Phi^2\right).
\end{equation}
According to the prescriptions of AdS/CFT correspondence~\cite{witten,gub} the five-dimensional mass $m_5$
of scalar fields is related with the canonical dimension $\Delta$ of corresponding scalar operators of dual
four-dimensional gauge theory living on AdS\(_5\) boundary as
\begin{equation}
\label{m5}
m_5^2R^2=\Delta(\Delta-4).
\end{equation}
The spectrum of normalizable modes of this model is (see, e.g.,~\cite{afonin2020})
\begin{equation}
\label{spSW}
  M^2_n=4|c|(n+\Delta/2),
\end{equation}
where $n=0,1,2,\dots$.

There are alternative formulations of SW model~\eqref{SW} which lead to the same classical equations of motion and
correlation functions. They are summarized and discussed in detail
in the recent work~\cite{Afonin:2021cwo}. For our purpose, we will use a formulation without dilaton background.
In general, if a SW model is given by the action
\begin{equation}
\label{Bz}
  S=\int d^4x dz\sqrt{g}\,B(z)\,\mathcal{L},
\end{equation}
where $B(z)$ is some $z$-dependent dilaton background,
one can redefine the fields so that the action~\eqref{Bz} takes the form
\begin{equation}
  S=\int d^4x dz\sqrt{\tilde{g}}\,\tilde{\mathcal{L}},
\end{equation}
with some modified metric. In the scalar case, the modified metric becomes~\cite{Afonin:2021cwo}
\begin{equation}
\label{tr}
  \tilde{g}_{MN}=B^{2/3}g_{MN}.
\end{equation}

The linear scalar spectrum~\eqref{spSW} can be generalized to
\begin{equation}
\label{spSW2}
  M^2_n=4|c|(n+\Delta/2+b),
\end{equation}
where $b$ is a free intercept parameter. The inclusion of this parameter into the scalar SW model
requires generalization of the dilaton background $e^{cz^2}$ to
$B=e^{cz^2}U^2(b,-1,cz^2)$, where $U(b,-1,cz^2)$ is
the Tricomi function~\cite{Afonin:2021cwo}.
The transformation~\eqref{tr} leads then to the following
modification of the metric~\eqref{metr},
\begin{equation}
\label{metr2}
  ds^2=f(z)\frac{R^2}{z^2}\lb\eta^{\mu\nu}dx_\mu dx_\nu-dz^2\rb,
\end{equation}
\begin{equation}
  f(z)=e^{2cz^2/3}U^{4/3}(b,-1,cz^2).
\end{equation}

The considered extension of holographic SW model by introducing the intercept parameter $b$ was first proposed in Ref.~\cite{Afonin:2012jn}.
The quantity $b$ parametrizes the fermion (namely quark) effects like the impact of chiral symmetry breaking
on the mass spectrum~\cite{Afonin:2021cwo}.

A direct holographic description of Higgs sector inevitably predicts an infinite tower of ``radially excited''
Higgs particles\footnote{A possible exception could consist in consideration of specific holographic models describing a finite number
of normalizable modes, such an example was constructed in Ref.~\cite{Afonin:2009xi}.}. This is definitely not what we want in our model.
The simplest way out is to motivate why the extra solutions are not physical.
Below we provide such a motivation.

The bottom-up holographic models lead to the two-point correlators in the form (in the momentum space)
\begin{equation}
\label{cor}
\left\langle O(q)O(-q)\right\rangle=\sum_{n=1}^\infty\frac{F_n^2}{q^2-M_n^2+i\varepsilon}.
\end{equation}
Exactly this form of correlators emerges in the large-$N$ limit of QCD~\cite{hoof,wit}, in which the
only singularities of the two-point correlation function of a hadron current operator $O$ are one-hadron states~\cite{wit}.
The large-$N$ scaling of appearing quantities is:
$M_n=\mathcal{O}(1)$ for masses,
$F_n^2=\langle0|O|n\rangle^2=\mathcal{O}(N)$ for residues,
$\Gamma=\mathcal{O}(1/N)$ for the full decay width~\cite{wit}. The last scale tells us that
the large-$N$ limit describes the zero-width approximation.
According to the principles of AdS/CFT correspondence~\cite{mald,witten,gub}, the large-$N$ limit
is necessary for building a holographic dual theory. In the practical holographic models,
the infinite sum over radially excited states in~\eqref{cor} is identified with the infinite tower
of 4D Kaluza-Klein-like excitations of a 5D field. This identification, however, has certain problems~\cite{Csaki:2008dt,afonin2020}.
A physically motivated way for description of the radially excited hadrons seems to consist in introducing operators
of higher dimensions which correspond to different 5D fields in dual theory~\cite{afonin2020}. This path (additional
higher dimensional operators) is used in QCD lattice calculations of radial hadron spectra. The particle states corresponding to $n>1$
in~\eqref{cor} should be then discarded as redundant solutions, more strictly, they should be replaced
by ``perturbative continuum''. Exactly this ansatz ``one infinitely narrow resonance + perturbative continuum''
has been successfully used in QCD sum rules since the late 1970s~\cite{rry}. The physical reason is that
the resonance width grows with mass and in practice the excited states overlap so strongly that often become
almost indistinguishable from the perturbative background. As the BSM field theory is expected to be more strongly coupled
than QCD, the resonances should be much wider in this theory, thus the approximation ``one resonance + smooth background''
should be even more justified.

Following the discussion above, we will consider only the ground state as
a physical one, other states corresponding to $n>1$ will be discarded as an artifact of the large-$N$ approximation.

\section{Confining behavior and breaking of effective string}

If the BSM strongly coupled theory is similar to QCD, it should lead to linearly growing with distance energy between two
static sources when the distance is large enough,
\begin{equation}
\label{pot}
  E\simeq\sigma r + \text{const},
\end{equation}
where \(\sigma\) is the tension of ``flux tube'' which is usually approximated by a string. In a pure Yang-Mills theory,
the strings are closed. If the fermions are introduced, a string between fermion and antifermion is open and
its tension is half of tension of the closed string. This property can be used for prediction of spectrum of composite states
since the mass spectrum is expressed (in a model-dependent way) via $\sigma$.

Suppose that the fermion effects can be parametrized by a parameter $b$ within some dynamical model. This will entail
the dependence of tension $\sigma$ on this parameter, $\sigma=\sigma(b)$. The physical value of $b$ is such that the
tension is halved in comparison with the case $b=0$ when the fermion effects are absent. This gives the equation for $b$,
\begin{equation}
\label{sigma_ratio}
  \sigma(b)=\frac{1}{2}\sigma(0).
\end{equation}

For a specific implementation of this idea we need a specific model. Such a model was proposed in~\cite{Afonin:2021zdu}
basing on the SW holographic approach~\cite{son2}. We will further follow the method of Ref.~\cite{Afonin:2021zdu}.

The holographic calculation of potential between two sources from the expectation value of the Wilson loop was suggested by
Maldacena~\cite{Maldacena:1998im}. Consider a rectangular Wilson loop located in the
4D boundary of the Euclidean AdS\(_5\) space.
In the limit of large Euclidean time, \(T\to\infty\), the expectation value of such Wilson loop is proportional to
\begin{equation}
\label{wl_ev_1}
  \left\langle W(\mathcal{C})\right\rangle\sim e^{-TE(r)},
\end{equation}
where \(E(r)\) corresponds to the energy of the fermion-antifermion pair. On the other hand,
it can be computed as~\cite{Maldacena:1998im}
\begin{equation}
\label{wl_ev_2}
  \left\langle W(\mathcal{C})\right\rangle\sim e^{-S},
\end{equation}
where \(S\) represents the area of a string world-sheet which produces the loop
\(\mathcal{C}\). The static energy follows from comparing~\eqref{wl_ev_1} and~\eqref{wl_ev_2},
$E=S/T$.

Within the bottom-up holographic QCD, the given idea was developed by
Andreev and Zakharov for the case of vector SW holographic model~\cite{Andreev:2006ct}.
The analysis is based on the Nambu-Goto string action,
\begin{equation}
\label{ng}
  S=\frac{1}{2\pi\alpha'}\int d^2\xi\sqrt{\det G_{MN}\partial_\alpha X^M\partial_\beta X^N},
\end{equation}
where \(g_{MN}\) is the modified AdS\(_5\) metric, $\alpha'$ is the tension of fundamental string. This action
is known to describe the area of string world-sheet. The expectation value of Wilson loop
can be calculated for this area and after that the dependence $E(r)$ can be extracted.
For the case of generalized scalar SW model, the corresponding calculation was done in Ref.~\cite{Afonin:2021zdu},
the final answer is given in a parametric form by the expressions
\begin{equation}
\label{dist}
  r=2\sqrt{\frac{\lambda}{c}}\int\limits_0^1dv\,
  \frac{U^{4/3}(b,-1,\lambda)}{U^{4/3}(b,-1,\lambda v^2)}
  \frac{v^2e^{2\lambda(1-v^2)/3}}
  {\sqrt{1-v^4e^{4\lambda(1-v^2)/3}\frac{U^{8/3}(b,-1,\lambda)}{U^{8/3}(b,-1,\lambda v^2)}}},
\end{equation}
\begin{equation}
\label{en}
  E=\frac{R^2}{\pi\alpha'}\sqrt{\frac{c}{\lambda}}\lsb\int\limits_0^1\frac{dv}{v^2}\lb
  \frac{e^{2\lambda v^2/3}U^{4/3}(b,-1,\lambda v^2)}{\sqrt{1-v^4e^{4\lambda(1-v^2)/3}\frac{U^{8/3}(b,-1,\lambda)}{U^{8/3}(b,-1,\lambda v^2)}}}-D\rb-D\rsb,
\end{equation}
where  \(D\equiv U^{4/3}(b,-1,0)\) is the regularization constant and
\begin{equation}
  z_0\equiv\left.z\right|_{y=0},\qquad
  v\equiv\frac{z}{z_0},\qquad
  \lambda\equiv cz_0^2.
\end{equation}
The rectangular Wilson loop in this calculation was parametrized by the coordinate choice
\(\xi_1=t\) and \(\xi_2=y\), where \(0\leq t\leq T\).

\begin{wrapfigure}{R}{0.3\textwidth}
  \vspace{-7mm}
  \begin{center}
    \includegraphics[width=0.3\textwidth]{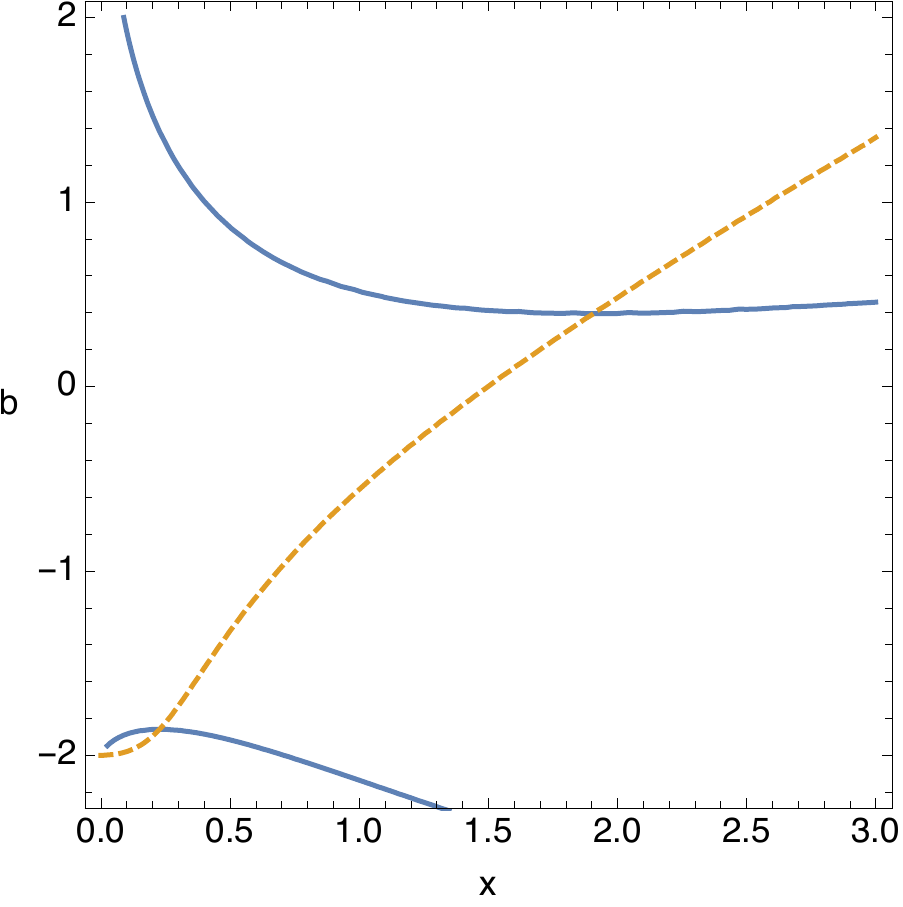}
  \end{center}
  \vspace*{-7mm}
  \caption{\footnotesize The conditions~\eqref{scalar_breaking} (blue solid) and~\eqref{realscalar} (orange
  dashed).}
  \vspace*{2mm}
  \label{sys_sols_scalar}
\end{wrapfigure}
Combining~\eqref{dist} and~\eqref{en}, the large distance asymptotics for the energy can be extracted~\cite{Afonin:2021zdu},
\begin{equation}
\label{en2}
  E\underset{r\to\infty}{\sim}\frac{R^2}{2\pi\alpha'}\frac{e^{2x/3}U^{4/3}(b,-1,x)}{x}cr.
\end{equation}
The parameter \(x\) is equal to one of the roots of derivative of the expression under the
square root in the integrals above.
The effective string tension between static sources as a function of intercept parameter $b$
follows from~\eqref{en2},
\begin{equation}
\label{stringscalar}
  \sigma(b)\equiv\frac{R^2}{2\pi\alpha'}\frac{e^{2x/3}U^{4/3}(b,-1,x)}{x}c.
\end{equation}
The condition~\eqref{sigma_ratio} for $\sigma(b)$ taken from~\eqref{stringscalar}
results in the equation,
\begin{equation}
\label{scalar_breaking}
      \dfrac{3e^{2x/3-1}U^{4/3}(b,-1,x)}{2x}=\dfrac{1}{2}.
\end{equation}

For consistency of the calculation, the expressions under the square roots in~\eqref{dist} and~\eqref{en}
must be non-negative in the whole interval of integration variable $v$.
This reality condition leads to the second equation,
\begin{equation}
\label{realscalar}
  1-\frac{2}{3}x-\frac{4}{3}x\frac{U'(b,-1,x)}{U(b,-1,x)}=0.
\end{equation}

We obtained thus two nonlinear equation for two unknown variables $x$ and $b$. The graphical solution is displayed in
Fig.~\ref{sys_sols_scalar}. The system has two solutions (the apparent intersection at $(x,b)=(0,-2)$
in Fig.~\ref{sys_sols_scalar} is not a solution to the Eq.~\eqref{scalar_breaking}), the corresponding numerical solutions
for the intercept parameter $b$ are
\begin{equation}
\label{scalar_intercept}
b_1\approx-1.859,\qquad  b_2\approx0.394.
\end{equation}

\section{Higgs masses}

According to our previous reasoning we associate the two solutions~\eqref{scalar_intercept} with two possible scalar spectra
in which only the ground states correspond to physically distinguishable particles. The operator describing
the coupling of Higgs field $h$ to a fermion-antifermion pair $\bar{\psi}\psi$ has the canonical mass
dimension $\Delta=4$. Within the framework of the proposed model, the relation~\eqref{spSW2} yields then
the following mass of the SM Higgs boson,
\begin{equation}
\label{h}
  M^2_h=4|c|\left(2+b_1\right).
\end{equation}
The mass of the second Higgs boson $h'$ is given in terms of~\eqref{scalar_intercept} and~\eqref{h} by
\begin{equation}
\label{h2}
  M^2_{h'}=\frac{2+b_2}{2+b_1}M^2_h.
\end{equation}
Substituting the corresponding numerical values from~\eqref{scalar_intercept} and $M_h=125$~GeV~\cite{pdg}
we get
\begin{equation}
\label{h2num}
  M_{h'}\approx515~\text{GeV}.
\end{equation}

A scalar particle with the given mass value was not observed. This is not surprising because a SM like Higgs boson
with a mass near  $500$~GeV should have the total decay width near $70$~GeV~\cite{spira}.
Such a broad resonance can be difficult to extract from the experimental background, especially taking into account
that $h'$ should be even much broader due to additional decays into the light higgs bosons.
If $h'$ couples to the SM only via mixing with the ordinary Higgs boson $h$ then one expects a considerable suppression
of $h'$ production cross section compared to that for the SM higgs boson, roughly about factor 10 according to the
analysis in Ref.~\cite{Godunov:2015nea}.
\begin{figure}[!b]
  \begin{center}
    \includegraphics[width=0.7\textwidth]{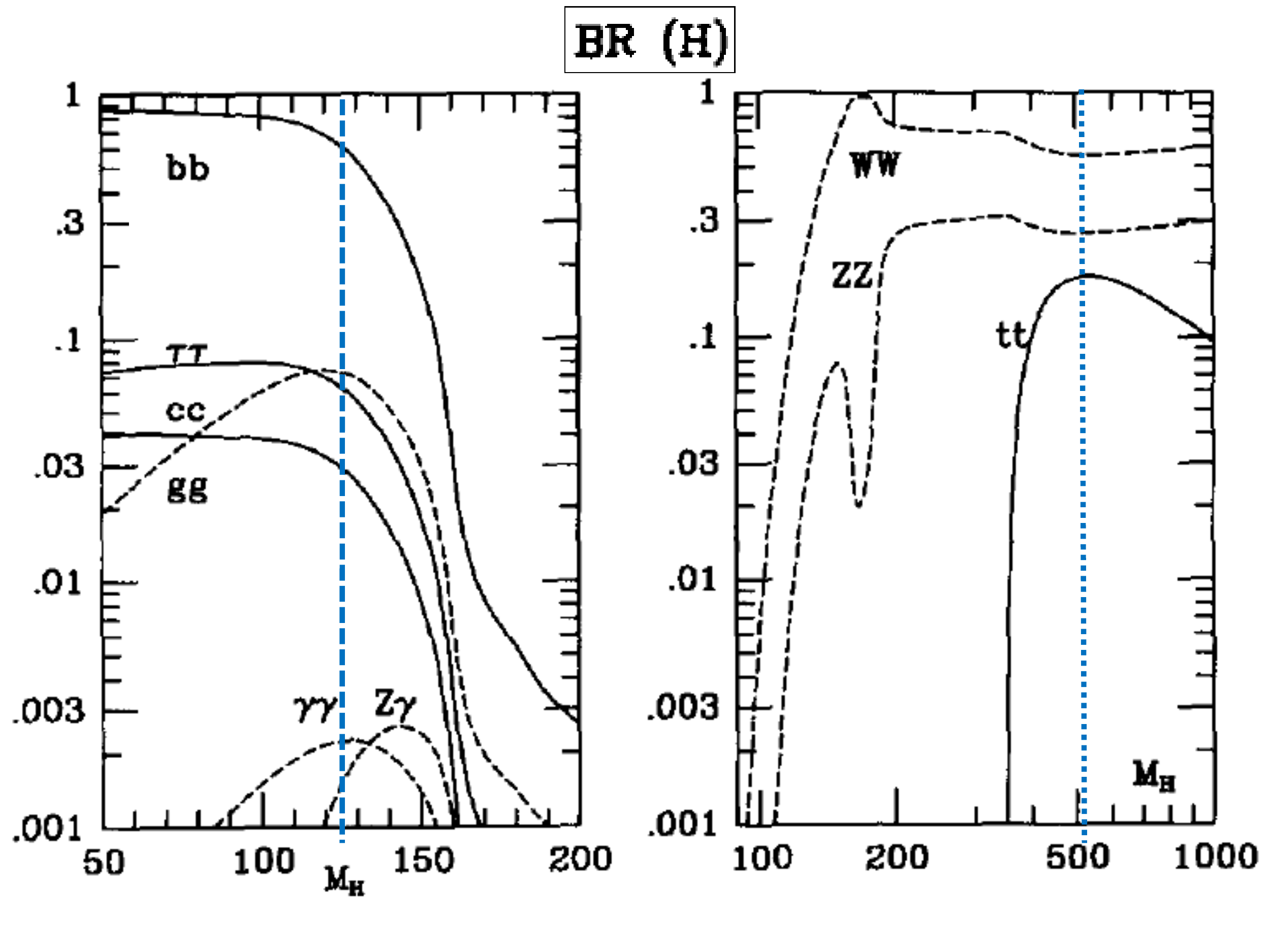}
  \end{center}
  \vspace{-6mm}
  \caption{\footnotesize The branching ratios of the dominant decay modes of SM Higgs boson as a function of its mass (in GeV).
  All known QCD and leading electroweak radiative corrections are included.
  The plot is taken from Ref.~\cite{spira}. The values of the SM Higgs mass and the predicted one~\eqref{h2num} are shown in blue
  (the dashed and dotted lines, correspondingly).}
  \label{branchings}
\end{figure}

On the other hand, our prediction should be considered as a prediction made in the large-$N$ limit of underlying $SU(N)$
gauge theory, i.e. deviations of the order of 20\% are acceptable. Several searches for additional Higgs bosons at the LHC
performed by the CMS and ATLAS Collaborations show an excess of events of about 3~$\sigma$ standard deviations above the
background expectation around a mass scale $M_{h'}\approx400$~GeV~\cite{Biekotter:2021qbc}. These Collaborations have
recently announced an even more pronounced excess of events (at a similar level of about 3~$\sigma$) near 650~GeV
with a total width of $\Gamma_{h'(650)}\approx100$~GeV which is compatible with the SM higgs particle of the given mass~\cite{Kundu:2022bpy}.
Our prediction~\eqref{h2num} lies close to the mean of these two observations. Also one can notice that the observed excess
$h'(650)$ is close to the threshold of simultaneous production of our predicted scalar $h'(515)$ and the standard higgs $h(125)$. Taking into account
the SM total width of $h'(650)$, this might suggest that $h'(650)$ could represent a kind of excitation of $h(125)$ via the
absorbtion of $h'(515)$. Definitely more data from the future Run of LHC are needed to check such an unusual scenario.

It is interesting to take a closer look at the predicted value~\eqref{h2num}.
We show in Fig.~\ref{branchings} the position of Higgs masses under consideration on the plot of Higgs branchings.
It is seen from the right part of Fig.~\ref{branchings} that the predicted mass value of~\eqref{h2num} corresponds to the maximal
$t\bar{t}$  branching ratio. In other words, it corresponds to the point in the region $M>M_h$ where the direct decay to fermion-antifermion
pair is most favorable. Even if the genuine second Higgs particle does not exist, we find the given holographic prediction
highly nontrivial.

It is curious to observe that the standard Higgs boson mass corresponds to the maximal branching ratio for decay into massless
gauge bosons, see the left part of Fig.~\ref{branchings}. Simultaneously, the branching ratios for decays into fermions begin to
rapidly decrease in the given region. We do not know a physical explanation but it is clear that the Higgs mass appears to be located
at the point where the production of Higgs boson via the gluon fusion (as in $pp$-collisions at CERN) is most favorable.
Within the presented model, one could speculate that the first solution describes the splitting into
two massless gauge bosons. The canonical mass dimension $\Delta=4$ refers then to the gauge-invariant field operator $G_{\mu\nu}^2$
and the energy $E(r)$ should be interpreted as a static energy between two ``colored'' sources.

This observation might shed light on the physical reason for the existence of two composite Higgs bosons:
One can construct two gauge and renormalization group invariant scalar operators in the standard gauge theories,
$O_1=\beta G_{\mu\nu}^2$, where $\beta$ is the Gell-Mann--Low beta-function, and $O_2=m_\psi\bar{\psi}\psi$. Both operators have the canonical dimension
$\Delta=4$. The two-point correlation functions $\Pi_1=\langle O_1O_1\rangle$ and $\Pi_2=\langle O_2O_2\rangle$ calculated in the background of BSM
strong sector may have poles. Then $M_h$ and $M_{h'}$ could arise form poles of different correlation functions, $\Pi_1$ and $\Pi_2$.

Finally it is curious to note that the numerically obtained relation $M_{h'}\approx4M_{h}$ is close to the approximate relation between
masses of first scalar excitations in QCD, $M_\sigma\approx4M_\pi$, where the broad scalar $\sigma$-meson is identified with the resonance
$f_0(500)$~\cite{pdg}.

\section{Conclusions}

Assuming the existence of confining ``strong sector'' beyond the Standard Model and that this sector can be described holographically,
we applied the Wilson loop analysis of confining behavior of generalized holographic Soft Wall model~\cite{Afonin:2021cwo} in the scalar
sector to description of the composite Higgs particles. The generalization consists in introducing the Regge ``intercept'' parameter
which, in a close analogy with the holographic description of standard strong interactions, parametrizes the fermion effects on the masses
of composite Higgs bosons. The constructed model predicts the mass of the second SM like Higgs boson near $515$~GeV. This prediction has an
unexpected and nontrivial feature --- the given value nearly coincides with the maximal branching ratio for decay into the $t\bar{t}$ pair.

Finally, a few words about prospects for future observation of the second Higgs particle.
At Run 2 LHC and especially in the future high-luminosity LHC studies, a smoking gun of a
heavy Higgs resonance is a strong resonant enhancement of double higgs
production in the $pp\rightarrow hh$ reaction~\cite{Godunov:2015nea,Barger:2014taa} ---
the corresponding cross-section should become several times larger than the SM value~\cite{Godunov:2015nea}.
The Higgs-singlet mixing should result in a substantial deviation of the extracted Higgs
trilinear self-coupling from its SM value due to a contribution of the process $pp\rightarrow h'\rightarrow hh$.
The mass of $h'$ and its mixing angle with $h$ may be extracted from a study of subsequent decays.
According to the analysis of Ref.~\cite{MartinLozano:2015vtq}, an efficient decay channel in a region of
masses $260<M_{h'}<500$~GeV is $hh\rightarrow b\bar{b}WW$, while for $M_{h'}>500$~GeV more efficient decay channel
is $hh\rightarrow b\bar{b}b\bar{b}$. Our predicted value of $M_{h'}\approx 515$~GeV lies near the border of these
two mass regions, so the both decay channels should be equally efficient for detecting possible signals
from the production of $h'$.

\section*{Acknowledgements}
I am grateful to Timofey Solomko for the joined work on the article~\cite{Afonin:2021zdu} which served as the basis for the present work.
This research was funded by the Russian Science Foundation grant number~21-12-00020.

\end{document}